# Critical evaluations of different implementations of Edwards volume ensemble

Houfei Yuan,[1] and Yujie Wang[1,2,3*]

[1]*School of Physics and Astronomy, Shanghai Jiao Tong University, Shanghai 200240, China*
[2]*School of Physics, Chengdu University of Technology, Chengdu 610059, China*
[3]*State Key Laboratory of Geohazard Prevention and Geoenvironment Protection, Chengdu University of Technology, Chengdu 610059, China*

**Abstract:** Granular systems can display reproducible microscopic distributions governed by a few macroscopic parameters, parallel to equilibrium statistical mechanics. Building on this analogy, Edwards's pioneering framework proposes a volume ensemble of equiprobable jammed states, introducing compactivity $\chi$ as an effective temperature. Despite its promise, debates persist regarding the framework's formulation and validity, with practical implementation often exposing inconsistencies. This study systematically examines different implementations using experimental data from spherical particle packings with varying friction coefficients, subjected to tapping and shearing. Our findings show that the heat capacity and overlapping histogram methods yield consistent $\chi$ values. Discrepancies in other approaches, such as the free volume model, primarily stem from differing reference-state compactivities and microstate definitions, whereas the mean-field theory, which effectively accounts for mechanical stability within a geometric coordination-number phase space, is constrained by its isostatic assumption and inherent approximations. Additionally, an effective temperature $\chi_f$ defined on Delaunay tetrahedra better describes topological excitations, but represents a quasi-equilibrium state, with true equilibrium achieved only under $\chi$. This study refines and expands our understanding of the Edwards volume ensemble in line with Edwards's original ideas, highlighting the combined contributions of friction and inherently disordered packing structures.

1. **Introduction**

Granular materials, which are inherently out of equilibrium, settle into mechanically metastable states when undisturbed. Under external driving, such as tapping or cyclic shearing, these systems can reach stationary states where the packing fraction stabilizes, independent of the preparation history[1-3].

Edwards and coworkers pioneered a statistical mechanical framework for jammed granular packings, treating volume as an analog of energy, with compactivity $\chi$ as the conjugate temperature-like variable[1, 4]. In this framework, the Hamiltonian is replaced by a volume function $W(\mathbf{q})$, which expresses the volume of the system as a function of its degrees of freedom $\mathbf{q}$. Granular entropy is then defined as $S(V) = \ln \Omega(V)$, where $\Omega(V) = \sum \delta(V - W(\mathbf{q}))\Theta(\mathbf{q})$ is the density of states (DOS) for a given volume $V$ under the mechanical stability condition $\Theta(\mathbf{q})$. In steady states, external driving induces fluctuations in the volume, similar to a thermal system in contact with a heat bath at fixed temperature. The probability of observing a microstate with a specific volume $V$ follows a Boltzmann-like distribution,

$$P(V) = \frac{\Omega(V)}{\mathbb{Z}(\chi)} \exp(-V/\chi), \tag{1}$$

where the partition function $\mathbb{Z}(\chi) = \int \Omega(V) \exp(-V/\chi)$. Building on this analogy to standard statistical mechanics, compactivity $\chi$, density of states $\Omega(V)$, and granular entropy $S(V)$ can be directly measured[5-10] . This framework has been instrumental in advancing the understanding of various granular properties, including segregation, critical state behavior, slow relaxation, plasticity, and structural defects[10-13]. Recent advancements in the Edwards statistical framework have further clarified the role of friction[8], demonstrated the equivalence between fluctuation-dissipation temperature and compactivity $\chi$ [14], and developed a new thermodynamic free energy formulation for granular materials based on Edwards' statistical mechanics framework[12].

Several methods have been developed to implement Edwards volume ensemble[5-7, 9, 15]. However, there are notable discrepancies between these methods. Understanding the theoretical foundation of each approach and the origins of these inconsistencies are critical. In addition, a significant challenge remains in the direct theoretical enumeration of mechanically stable configurations. Experimental techniques, such as heat capacity and overlap histogram methods, can explicitly measure the density of states $\Omega(V)$ [9]. However, these methods lack the capability to build theoretical models that explicitly describe $\Omega(V)$. Although many other theoretical models have been proposed to derive $\Omega(V)$ directly[15-18], these models often produce results that conflict with experimental observations[6]. Such inconsistencies suggest that certain aspects of these models may not accurately reflect the physical realities of granular systems. Although recent advancements in the Edwards ensemble primarily determine the volume function based on Voronoi tessellation, volume statistics based on Delaunay tetrahedra also exhibit certain quasi-equilibrium behaviors that differ from those based on Voronoi cells[19]. Understanding the relationship between the volume function associated with the Voronoi cells and Delaunay tetrahedra is essential for developing a more unified theoretical framework.

In this work, we critically evaluate the various models proposed to implement the Edwards volume ensemble and calculate $\Omega(V)$ using our experimental data obtained from monodisperse spherical packings with various friction coefficients, prepared using tapping and shearing methods[8, 13]. Furthermore, we compare Delaunay and Voronoi tessellations, highlighting Delaunay tessellation's superior capability in capturing detailed topological structures and their fluctuations in granular systems.

## 2. Measuring the Edwards compactivity $\mathcal{X}$

Empirical observations have revealed that mechanically stable sphere packings can only form within a finite range of packing fractions[20-22]. This range of disordered configurations is bounded by Random Loose Packing (RLP) at the lower limit and Random Close Packing (RCP) at the upper limit[23-26]. Building on this understanding, the canonical Edwards volume ensemble provides a theoretical approach for describing such packings. Various methods have been developed to implement this concept in both experimental and simulation systems[5-7, 9, 15, 17]. A key objective of these implementations is the measurement of compactivity $\mathcal{X}$, which, in principle, should yield consistent results if all methods accurately adhere to Edwards's conjecture.

The volume fluctuation method, also known as the specific heat method, is analogous to the relationship between energy fluctuations and the specific heat in thermally equilibrated systems. In this approach, $\mathcal{X}$ can be determined from intensive volume fluctuation: $\mathrm{var}(\phi) = \sigma_V^2 / m$, where $\sigma_V^2 = \langle V^2 \rangle - \langle V \rangle^2$ is the variance of the volume $V$, calculated over a coarse-grained region containing $m$ particles to eliminate correlations among neighboring

particles. Using the RLP state as a reference, $\chi$ can be obtained as:

$$\frac{1}{\chi(\phi)} - \frac{1}{\chi(\phi_{RLP})} = \int_{\phi_{RLP}}^{\phi} \frac{d\phi}{\phi^2 \operatorname{var}(\phi)} . \tag{2}$$

It is clear that $\chi$ can be uniquely determined when $\chi(\phi_{RLP})$ can be defined.

The overlapping histogram method offers an alternative approach for measuring compactivity $\chi$. It examines the ratio of volume probability distributions between two states prepared at different $\chi$ for the same system, which, according to Eq. (1), should follow an exponential dependence on $V$. Using again the RLP state as a reference, the ratio can be expressed as:

$$\frac{P(V, \chi_{RLP})}{P(V, \chi)} = \frac{\mathbb{Z}(\chi)}{\mathbb{Z}(\chi_{RLP})} e^{\left(\frac{1}{\chi} - \frac{1}{\chi_{RLP}}\right)V} . \tag{3}$$

By performing a linear fit of the logarithmic ratio of the experimentally obtained volume probability distributions $\ln \frac{P(V, \chi_{RLP})}{P(V, \chi)}$ as a function of $V$, $\chi$ can be determined up to an undetermined $\chi(\phi_{RLP})$.

When applied to the volume analysis of Voronoi tessellation in both 2D and 3D systems, both the heat capacity method and the histogram method exhibit excellent consistency [6, 8, 27]. This agreement is expected because both methods are derived from Eq. (1). However, the overlapping histogram method offers a significant advantage in that it provides a direct test of the Boltzmann distribution assumption. Specifically, the logarithm of the ratio of volume probability distributions is required to depend linearly on $V$, a condition experimentally confirmed for all pairs of volume distributions measured. In contrast, the volume fluctuation method cannot be directly employed to test the Boltzmann form, as it can yield $\chi$ even when the Boltzmann distribution is not strictly followed, provided that fluctuations exist in the system.

Despite their applicability, both the heat capacity method and histogram overlapping method share a fundamental limitation: they do not directly determine the compactivity $\chi$ of the reference state. This is because both the density of states $\Omega(V)$ and Boltzmann factor $\exp(-V/\chi)$ are functions of the volume $V$, making it impossible to extract the absolute value of $\chi$. Traditionally, the RLP state, $\phi_{RLP} \approx 0.56$, which corresponds to the loosest packing achievable for systems with an infinite friction coefficient, has been assigned infinite compactivity[7]. This assignment is based on the assumption that the entropy reaches its maximum at the RLP state[17, 28]. Alternatively, the friction-dependent RLP has also been proposed, representing the loosest packing achievable for a given system with a specific finite friction coefficient $\mu$ [5, 17]. In this scenario, $\chi(\phi_{RLP}) = \infty$ occurs at different $\phi_{RLP}$ for different friction systems. Experimental studies have shown that compactivity, as defined by this friction-dependent RLP, agrees with the fluctuation-dissipation temperature derived from transport measurements[14]. Furthermore, a granular analog of the thermodynamic zeroth law emerges only when the loosest packing state for each friction coefficient is specified as the state with infinite compactivity [8, 13, 14]. These observations highlight the critical role of appropriately specifying the reference state when implementing the Edwards volume ensemble.

Although the heat capacity and histogram overlapping methods successfully implement the Edwards volume ensemble and provide practical ways to measure compactivity, they fail to offer deeper physical insights. For instance, it remains unclear why packings with different friction coefficients exhibit similar volume distributions at the same packing fraction[8]. Addressing this limitation requires the development of refined microscopic theories. We also note that these methods primarily determine the volume function using the Voronoi tessellation.

Alternative tessellation methods, such as the quadrons method and tensorial formulations[29, 30], have been proposed to partition space into non-overlapping regions to express the volume function. However, owing to practical challenges and inherent limitations, particularly in 3D systems, these methods are less commonly employed for measuring volume statistics of Edwards ensemble[4].

3. **The free volume model on the k-Gamma distribution**

Aste and Di Matteo proposed a free volume model to compute compactivity $\mathcal{X}$, which explicitly determines the density of state[15]. The free volume is defined as $V_{f,voro} = V - V_{min}$, where $V_{min}$ denotes minimum attainable volume. Given that experimentally measured Voronoi volume distributions are well-fitted by gamma distributions, they proposed a rescaled k-Gamma distribution,

$$P(V_{f,voro}) = \left(\frac{k}{\langle V_{f,voro} \rangle}\right)^k \frac{V_{f,voro}^{k-1}}{\Gamma(k)} \exp\left(-k \frac{V_{f,voro}}{\langle V_{f,voro} \rangle}\right), \tag{4}$$

where $\langle V_{f,voro} \rangle = \langle V \rangle - V_{min}$ is the mean free volume, and $k$ is the shape factor. Using this formulation, they obtained compactivity $\mathcal{X}_\Gamma = \frac{\langle V_{f,voro} \rangle}{k}$ and volume variance $\sigma_V^2 = \frac{\langle V_{f,voro} \rangle^2}{k}$. The density of state $\Omega(V_{f,voro})$ is proportional to $\frac{V_{f,voro}^{k-1}}{(k-1)!}$. Notably, this model does not require the microstates to be jammed and can be extended to the liquid regime.

Despite the apparent differences, Aste's free volume model is fundamentally aligned with the specific heat method, as both are derived from Eq. (1), which is the canonical Edwards ensemble expression for volume. In the granular jammed state, assuming ergodicity within the

Edwards volume ensemble, the density of state $\Omega(V_{f,voro})$ should be material-dependent but independent of the packing fraction $\phi$ or compactivity $\chi_\Gamma$. This suggests that the parameter $k$ should remain constant, consistent with experimental observations showing only narrow variations in $k$ in jammed packings ranging from RLP and RCP. Consequently, all distributions can be rescaled onto a master k-gamma function with a universal shape factor $k = 13.27$, as shown in Fig. 1 (b).

The primary distinction between Aste's free volume model and the heat capacity method lies in the choice of the reference state for infinite compactivity: the heat capacity method uses the RLP state, whereas the free volume model adopts the ideal gas state. This difference leads to a numerical shift in the inverse compactivities determined using two methods: $1/\chi_\Gamma(\phi) - 1/\chi_\Gamma(\phi_{RLP}) = 1/\chi(\phi)$. This shift arises from differences in how microstates are accounted for— the heat capacity method follows the original Edwards assumption, considering only mechanically stable states, whereas Aste's free volume model includes all microstates corresponding to a given volume, without imposing mechanical stability constraints.

Similar to traditional free volume theories (e.g., Turnbull's model), Aste defined free volume as the space in which a particle can move freely[31-34]. However, in granular packings, most particles are constrained by contact, leaving only a small fraction as rattlers, *i.e.* particles can freely move within the packing[35, 36]. Consequently, volume fluctuations in granular systems are purely configurational and arise from variations in the volumes of mechanically stable packings. This is in contrast to liquids, where the free volume originates from vibrational contributions[37, 38]. In Aste's model, the entropy derived from the free volume includes both configurational and vibrational contributions, similar to pore-size entropy[39]. As a result, the number of accessible microstates is overestimated when considering granular packing from RLP to RCP. This partly explains why the free volume model exhibits significant $k$ fluctuations below the RLP threshold[15]: when $\phi > 0.55$, only jammed configurations are considered and the Edwards volume ensemble remains valid; while $\phi < 0.55$, the jamming constraint is lifted, all configurations are included, and the Edwards ergodicity hypothesis is no longer expected to hold. This also explains why Aste's model does not predict any significant anomaly in the RLP state, as its entropy includes both configurational and vibrational contributions, which continue across the packing fraction of RLP.

## 4. The mean-field theory for frictional spherical systems

The full Edwards ensemble characterizes granular packing through both macroscopic volume and stress[4]. However, owing to the complexity of the analytical treatments, practical approaches often employ approximations. A key development in this direction is the mean-field geometrical framework developed by Makse and coworkers, which explicitly incorporates mechanical stability constraint into the partition function for frictional spherical granular packings[5, 17, 18]. This framework, based on the isostatic conjecture, employs mean-field calculations of the microscopic volume function using the geometric coordination number, and obtains mechanically stable packings by discretizing the geometric contact phase space. Through this approach, the density of states $\Omega(V) = \sum \delta(V - W(\mathbf{q}))\Theta(\mathbf{q})$ and partition function are derived, providing a simplified yet effective analytical representation.

Specifically, the geometric coordination number $z$ is employed to describe the packing structure and establish a direct relationship between volume $V$ and $z$. The system is assumed to be ergodic within the phase space defined by $z$. Additionally, the isostatic mechanical coordination number $Z(\mu)$ is introduced to specify the allowed range of $z$. For frictionless spherical particles, it is well established that the minimum average mechanical coordination number $Z$ required for mechanical stability is given by $Z = 2d = 6$, where $d$ is the spatial dimension[40, 41]. In the presence of friction ($\mu > 0$), fewer contacts are required for stability, with $Z \geq d + 1 = 4$ and $Z = 4$ for infinitely rough spheres[22]. $Z(\mu)$ varies continuously from four to six as $\mu$ decreases from infinity to zero.

By counting the states within this geometric contact phase space, the density of states can be reformulated as an integral, with its lower limit defined by $Z(\mu)$, expressed as

$$\Omega(V) = \int_{Z}^{6} P(V|z) g(z) dz,$$ where $g(z) \propto h_z^z$. Here, $h_z$ is a constant related to the spacing between the states in the contact phase space. In this framework, friction restricts the range of $z$ available for integration, limiting accessible states to those with $6 > z > Z_{RLP}$ ($\phi > \phi_{RLP}$).

Figure 2 (a) shows the phase diagram derived from this framework. Because the theory assumes an isostatic condition, the average mechanical coordination number $Z(\mu)$ remains essentially constant and is unaffected by variations in compactivity $\mathcal{X}$ or packing fraction $\phi$. The average geometric coordination number $z$ is solely a function of the packing fraction $\phi$, following the relationship $\phi = \dfrac{z}{z + 2\sqrt{3}}$, which is obtained through mean-field simulations of Voronoi cell volumes. Consequently, this mean-field theory predicts a phase diagram bounded by the RLP and RCP lines as well as a G-line.

Recent X-ray tomography experiments (Fig. 2(b)) have allowed for a direct comparison of this phase diagram by plotting $\phi$ as a function of $z$ for various systems[8]. Notably, strong agreements are observed among the experiments, numerical simulations, and mean-field theory at the RLP line. This agreement likely reflects the fact that, at RLP, the system is isostatic, with $z$ reaching its lower bound, so that $z = Z$. As a result, the RLP line for systems with different friction coefficients aligns with the $z - \phi$ relationship, given by $\phi = \dfrac{Z}{Z + 2\sqrt{3}}$.

However, this theory has several limitations, most notably its assumption of isostaticity, which is inconsistent with experimental observations. Experimental measurements of the geometrical coordination number $z$ show significant deviations from the theoretical relationship $\phi = \dfrac{z}{z + 2\sqrt{3}}$ away from the RLP line and exhibit a clear dependence on friction $\mu$. This contrasts with mean-field theory, which assumes a one-to-one correspondence between $\phi$ and $z$, with friction only defining the minimum $z$ for accessible states.

In practice, distinguishing between geometrical and mechanical contacts is often challenging, and the distinction may be more of a conceptual approximation than an experimental one. Granular packings are typically hyperstatic, meaning that particles have more contacts than strictly required for mechanical equilibrium. This contrasts with mean-field theory's assumption that packings are restricted to isostatic states. Consequently, the strict application of the isostatic condition can lead to discrepancies between theory and experiments, as experimentally measured $z$ generally falls between the isostatic coordination number $Z$ and theoretical $z$ predicted by mean-field theory. Because no universal one-to-one correspondence exists between $\phi$ and $z$ across systems with different friction coefficients, it is unsurprising that the experimental results do not yield a simple master curve for the density of states as a function of volume $V$.

Another limitation associated with mean-field theory is the treatment of configurational entropy. Figure 3 shows the experimentally measured entropy as a function of $z$. For tapped systems, the data exhibit a clear linear dependence, collapsing onto a single master curve that is well described by the linear relation, $\Delta S = k_z \Delta z$, with $k_z = 0.455$. Here, $k_z$ is analogous to the Boltzmann constant; however, its precise physical interpretation remains unclear. While the mean-field theory appears to accurately describe this relationship for tapped systems, it breaks down for sheared systems. For sheared systems, the $S - z$ relationship exhibits two-stage behavior: at large $\phi$, the system shows a linear $S - z$ dependence, similar to tapped systems, but with a smaller slope; at smaller $\phi$, the slope steepens to resemble that of tapped systems. To understand the origin of these discrepancies between tapping and sheared systems, as well as their deviation from mean-field theory, we employ a landscape perspective, as illustrated in Fig. 4. Both mean-field theory and our framework treat granular free energy landscapes as rugged energy landscapes, where each local minimum corresponds to a mechanically stable state. However, these two approaches differ significantly on the global scale. From our perspective, the granular free-energy landscape resembles a multi-basin glassy landscape, capturing the complex, disordered nature of granular systems[10]. In contrast, mean-field theory operates with a single basin, inherently neglecting complex structural correlations beyond first-shell neighbors. This omission is critical, as the presence of a glass-like landscape is essential for understanding slow compaction dynamics and the emergence of RLP and RCP states[42]. To emphasize this point, we note that our recent study reveals that the configurational entropy of granular packings closely mirrors the complexity observed in thermal glassy hard-sphere systems, particularly when analyzed as a function of the packing fraction $\phi$ [13]. This correspondence arises naturally when considering that jammed packings inherit configurations from their corresponding hard sphere liquids. The difference in the $S-z$ behaviors between tapped and sheared systems can be attributed to how these protocols explore the energy landscape. Tapping enables a more uniform exploration of the entire landscape by effectively elevating the system above the landscape during tapping, before quenching it into mechanically stable states. In contrast, at a large $\phi$ with a small driving amplitude, shear protocols perturb the system more locally, confining exploration in the vicinity of a single basin. Consequently, sheared systems only access mechanically stable configurations within a single basin of the disordered landscape, limiting their ability to explore other basins of the global energy landscape. The difference in the methods of different protocols in exploring the landscape leads to different $S-z$ behaviors. The smaller slope observed for the sheared system at large $\phi$ suggests that it is less effective at accessing the configurational entropy associated with the disordered landscape, akin to the limitations of the mean-field theory. In contrast, at small $\phi$, the sheared system can access both the entropy within a single basin and that of the broader disordered landscape, leading to an increase in slope. This behavior eventually becomes more aligned with the tapping protocol, which is more efficient for the exploration of the entire landscape under the quenching procedure.

Lastly, we emphasize that friction plays a more intricate role in real granular packings as compared to the simplified one as a limiting threshold in mean-field theory. In real granular systems, friction not only sets the threshold for the RLP state but also directly influences the mechanical stability of accessible states. This highlights the multifaceted effect of friction on both the structural and mechanical properties of granular packings, underscoring its critical role beyond the assumptions of the mean-field model.

5. **The topological structural characterization and the effective temperature $\chi_f$**

Delaunay tessellation is also widely used to analyze the volume statistics of disordered sphere packings. Unlike Voronoi tessellation, Delaunay tetrahedra are not uniquely associated with individual particles, making it challenging to decompose tessellation into single-particle contributions, thereby limiting its use for statistical analysis[4]. Nevertheless, Delaunay tessellation excels in capturing local topological structures and their evolution, particularly during plastic events, known as T1 events[43, 44].

The local topology of granular systems can be characterized by N-ring structures, which are groups of $N$ Delaunay tetrahedra that share a common edge[45]. These structures play a crucial role in understanding the structural and mechanical behaviors of granular materials. Notably, 5-rings are associated with glass-like order due to their formation from quasiregular tetrahedra, while other N-rings incorporate disclination defects[46]. Our previous study has identified 4-rings as the microscopic carriers of plasticity. Building upon this insight, the macroscopic constitutive relation can be constructed by summing the contributions of these plastic carriers under shear, following Langer's Shear Transformation Zone (STZ) theory[47].

However, it has been observed that when constructing the constitutive relation, the excitation of N-ring structures is more accurately characterized by an effective temperature $\chi_f$ rather than Edwards compactivity $\chi$. This presents a conundrum, as Edwards compactivity has been repeatedly validated in jammed granular systems. To resolve this discrepancy, we investigate the statistical behavior of N-ring structures under both $\chi_f$ and $\chi$.

To calculate $\chi_f$, we define the free volume of Delaunay tetrahedra as $v_f = v_t - v_{t,g}$,

where $v_t$ is the tetrahedron's volume, and $v_{t,g}$ is the particle volume within it[19]. The probability distribution function (PDF) of tetrahedral free volume, shown in Fig. 5(a), exhibits an exponential tail, indicative of an underlying Boltzmann-like distribution. Theoretical models treat Delaunay tetrahedra as structural building blocks capable of exchanging free volume[15, 16, 31]. The emergence of the exponential tail originates from entropy maximization, with the slope of the distribution defining the effective temperature $\chi_f$. It is worth noting that $\chi_f$ alone cannot capture all physics of the packing structure, as PDFs deviate from exponential behavior at small volumes, which remain largely unaffected by packing density, as shown in Fig. 5(a). These small tetrahedra arise due to steric constraints and particle friction.

Similar to $\chi$, $\chi_f$ depends on both $\phi$ and $\mu$, as illustrated in Fig. 5(b). Consequently, $\chi_f$ is incompatible with $\chi_\Gamma$ derived from the k-gamma distribution ($\chi_\Gamma = \frac{\langle V_{f,voro} \rangle}{k}$). Nonetheless, for all systems, $\chi_f$ decreases with $\phi$ and converges to a similar value at RCP, similar to $\chi$.

In our earlier study on cyclically sheared granular materials, we found that the effective temperature $\chi_f$ [19], plays a critical role in governing the evolution of N-ring structures and topological rearrangements. Analogous to thermal excitation of point defects in crystals, we hypothesize that the concentrations of different N-ring structures are governed by $\chi_f$ and follow a Boltzmann distribution based on their corresponding excitation volumes. Specifically, the ratios $N_{N-ring}/N_{5-ring}$ and $-1/\chi_f$ are expected to satisfy the following:

$$N_{N-ring}/N_{5-ring} \sim \exp(-\frac{V_{N-ring}}{\chi_f}), \quad (1)$$

where $N_{N-ring}$ represents the number of respective N-ring structures, and $V_{N-ring}$ denotes the mean excitation volume from the 5-ring ground state within these systems. In Fig. 6(a)-(c), we present the relationship $N_{N-ring}/N_{5-ring}$ as a function of $-1/\chi_f$ on a semi-logarithmic plot. Remarkably, all relationships exhibit clear linearity, consistent with the expected characteristics of the Boltzmann distribution and the thermal excitation scenario, even at low $\phi$. The excitation volume $E_{N-ring}$ for each type of N-ring structure is determined by fitting the slopes of the curves. The resulting excitation volumes are presented in Table I, with the units expressed in $v_g$. Surprisingly, despite the variations in $\phi$ and $\chi_f$ ranges, the excitation volumes exhibit remarkable consistency. The average excitation volumes for 3- to 7-ring structures are 0.040, 0.020, 0.008, 0.041 $v_g$, respectively. Notably, the 3-ring and 7-ring structures exhibit significantly larger excitation volumes compared to the 4-ring and 6-ring structures, reinforcing the notion of 5-ring structures as the ground state, with other N-ring structures serving as topological defects. Since these topological structures represent different basins in the free energy landscape, the consistency of excitation volume across systems with varying friction coefficients suggests that friction uniformly increases the average height of different N-ring energy basins.

Interestingly, $\chi_f$ also plays a crucial role in determining the distribution of N-ring structures across different systems. Figure 7(a) plots the concentration of 5-ring structures $c(N_{5-ring})$ as a function of $\chi_f$ for various systems. Remarkably, despite differences in friction coefficients $\mu$ and preparation protocols, $c(N_{5-ring})$ collapses onto a master curve. In Fig. 7(b), we compare the distributions of 3- to 7-ring structures for packings with similar $\phi$ and $\chi_f$. Interestingly, packings with similar $\chi_f$ exhibit more consistent ring distributions than those with similar $\phi$, underscoring the strong correlation between $\chi_f$ and N-ring topology.

This suggests an unexpected one-to-one correspondence between $\chi_f$ and the formation of N-ring structures, which extends beyond specific systems.

To further elucidate the role of $\chi_f$, we use the polytetrahedral order parameter $\delta$ to quantify the shape deviations from perfect local tetrahedron, defined as $\delta = l_{max}/d - 1$, where $l_{max}$ is the length of the longest edge of the tetrahedron[48, 49]. Figure 7(c) shows the $\langle\delta\rangle$, the average value of $\delta$, as a function of $\chi_f$. As $\chi_f$ decreases (corresponding to higher density), $\langle\delta\rangle$ decrease monotonically, collapsing onto a master curve across all systems, indicating that shape deviations are also well captured by $\chi_f$. A similar trend is observed when the shape deviations of Voronoi cells are characterized using their anisotropy index $\beta_{local}$, derived from the rank-2 Minkowski tensor $W_0^{0,2}$ [50]. $W_0^{0,2}$ is defined as $W_0^{0,2} = \int r \otimes r \, dV$, where $r$ is the vector from the particle center to each volume element. The anisotropy index $\beta_{local}$, calculated as the ratio of the maximum to minimum eigenvalues of $W_0^{0,2}$, is averaged over all particles to yield $\langle\beta_{local}\rangle$. Figure 7(d) plots $\langle\beta_{local}\rangle$ as a function of $\chi_f$, demonstrating collapse behavior similar to that of $\langle\delta\rangle$. This indicates that shape deviations of both the Delaunay tetrahedra and Voronoi cells are consistently represented by $\chi_f$.

The analysis suggests that $\chi_f$ is powerful for capturing the excitation of topological defects and local cell shape fluctuations in granular systems. This behavior closely links $\chi_f$ to the effective temperature associated with the configurational entropy in thermal disordered systems. However, because $\chi_f$ is not directly related to volume distributions at small volumes, it is less sensitive to particle contact and, therefore, the mechanical stability constraints inherent to granular systems. This distinction sets $\chi_f$ apart from Edwards compactivity $\chi$, as the latter simultaneously accounts for both the disordered structures and mechanical stability constraints in granular systems. Figure 8 illustrates the inverse of the two effective temperatures as a function of $\phi$. At low $\phi$, the evolution of $\chi^{-1}$ and $\chi_f^{-1}$ align for both tapping and shearing systems, reflecting a rough consistency between the two perspectives at high compactivity. However, at higher $\phi$, a divergence emerges: $\chi_f^{-1}$ evolves more slowly with increasing $\phi$, whereas $\chi^{-1}$ rises sharply. This discrepancy at high $\phi$ leads to deviations in the compactivity $\chi$ from the Boltzmann relation in Eq. (1), as illustrated in Fig. 6(d)–(f). The slower evolution of $\chi_f^{-1}$ suggests that topological changes, defined by Delaunay tessellations, become less pronounced at higher $\phi$. At these dense states, compaction predominantly arises from localized rearrangements within the same volume landscape basin, with minimal topological variations. These findings therefore resolve the apparent conflict between the two statistical frameworks, suggesting that $\chi_f$ is better suited for analyzing topological structures. In some sense, although the system can be considered as equilibrated in $\chi$ according to the Edwards ensemble, it can also be considered as quasi-equilibrated in $\chi_f$ when analyzing topological structures or local cell shape fluctuations, similar to the effective temperature of thermal hard-sphere systems.

To better understand the differences in entropy maximization with and without mechanical constraints, we examine the distinct behaviors of the tapping and shearing protocols at very low compactivity. For the tapped system, a kink appears around $\chi^{-1} \sim 7$, after which the increase in $\chi_f^{-1}$ slows down. In contrast, for the sheared system, kinks occur near $\chi^{-1} \sim 12$, followed by a plateau in $\chi_f^{-1}$ with a plateau value of 13. A potential explanation for the differing behaviors is that, at high compactivity, cyclic shearing is more effective in inducing topological changes as compared to tapping when varying $\phi$. The anisotropic nature of shearing allows it to explore the energy landscape more efficiently, overcoming free energy barriers between basins. By contrast, the isotropic nature of tapping limits its ability to facilitate such transitions, resulting in a slower increase in $\chi_f^{-1}$ as the systems become more compact. This difference again highlights the distinct ways in which different protocols navigate the energy landscape and induce structural changes other than their different impacts on $z$, as discussed above.

## 6. Conclusion

In this study, using experimentally obtained granular packing structures with varying friction coefficients prepared under tapping and shearing protocols, we evaluated several methods originally proposed to implement the Edwards volume ensemble to assess their consistency with one another and Edwards's original conjectures. We found that the heat capacity and overlapping histogram methods are in agreement with each other and more faithfully reflect Edwards' original ideas. Discrepancies with other methods primarily arise from differences in how the compactivity of the reference RLP state is specified, and whether the mechanical stability criterion is strictly enforced when counting microstates.

In addition, we critically assessed the mean-field theory proposed by Makse and coworkers. While the theory successfully translates the mechanical stability criterion into state counting within the geometrical coordination number phase space, it exhibits limitations in both the volume function and entropy calculation owing to the isostatic conjecture and its inherent mean-field nature.

We also address an intriguing observation regarding the effective temperature $\chi_f$ defined for Delaunay tetrahedra. This effective temperature $\chi_f$ appears to be more effective in capturing the excitation of topological structural defects and characterizing local cell fluctuations compared to Edwards compactivity $\chi$. We found that this discrepancy originates primarily from the way the mechanical stability criterion is implemented: $\chi_f$ is essentially the free volume effective temperature of disordered materials, which does not incorporate the mechanical stability criterion. As a result, it is better suited for capturing topological and local cell shape fluctuations. However, it is important to note that while the system is in quasi-equilibrium under $\chi_f$, it reaches strict equilibrium only under Edwards compactivity $\chi$, as Edwards originally conjectured.

**Conflicts of interest** There are no conflicts to declare.

**Acknowledgements** This work is supported by the National Natural Science Foundation of China (No. 12274292).

**Data availability statement** All data needed to evaluate the conclusions are presented in the paper. Raw data and corresponding experiment data are available upon reasonable request.

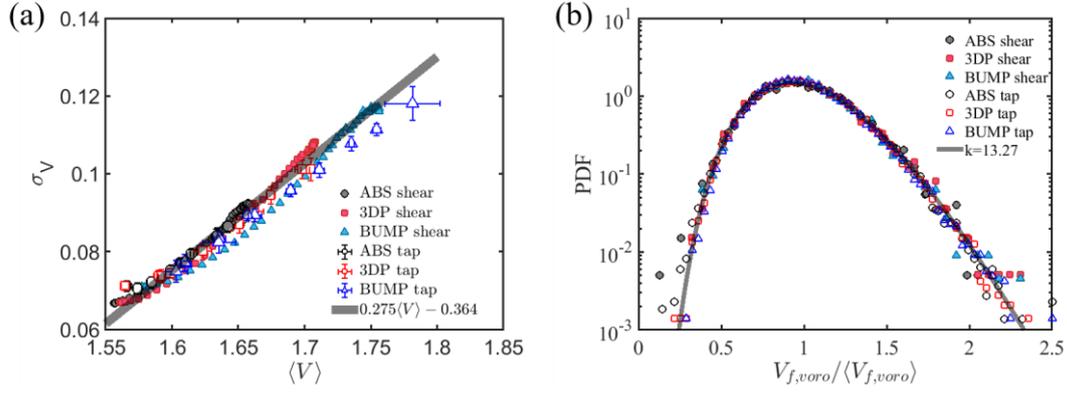

FIG. 1. (a) Voronoi volume fluctuation, $\sigma_V$, as a function of $\langle V \rangle$ for various systems. The gray line represents the linear fit of $0.275\langle V \rangle - 0.364$. (b) Voronoi volume distributions normalized by $\dfrac{V_{f,voro}}{\langle V_{f,voro} \rangle}$. All the distribution are well described by a k-Gamma distribution with $k = 1/0.275^2 = 13.27$.

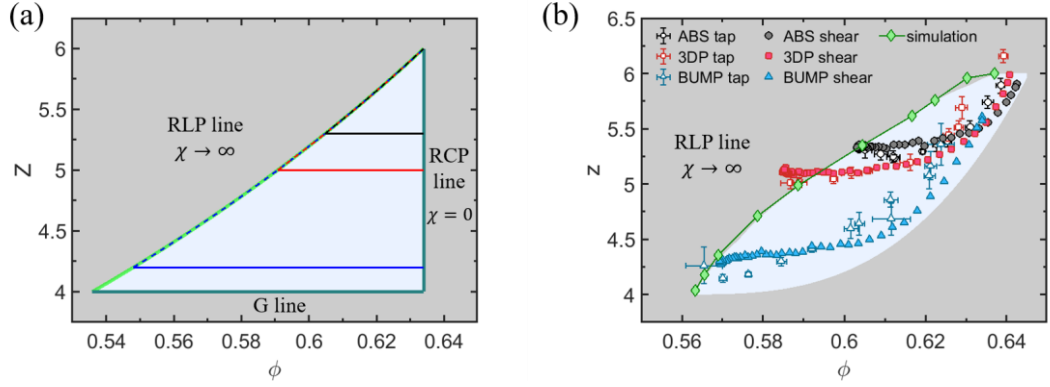

FIG. 2. $Z-\phi$ phase diagram of jammed states for frictional spherical systems obtained from (a) the mean-field theoretical framework developed by Song et al. and (b) X-ray tomography experiments. In panel (a), three iso-friction lines corresponding to experimental systems with specific friction coefficients are shown. Stable random jammed states are prohibited in the gray area. In the theoretical phase diagram, the iso-friction lines are approximately flat, indicating that the mechanical coordination number $Z$ remains nearly constant for a given friction coefficient. The geometrical coordination number $z$, as a function of $\phi$, follows the RLP line (dashed lines in panel (a)) and converges with the iso-friction line at the RLP state.

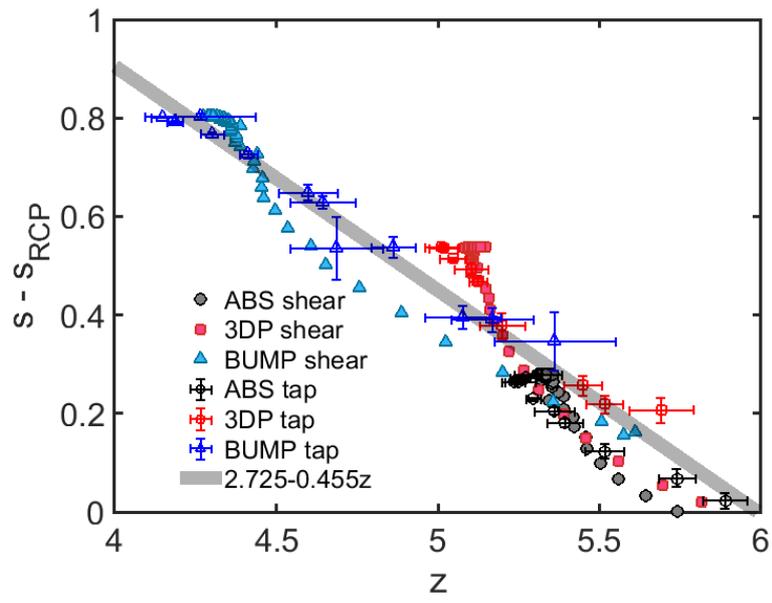

FIG. 3. Specific entropy $s$ as a function of coordination number $z$ for various systems. The gray line represents the linear fit for tapping systems.

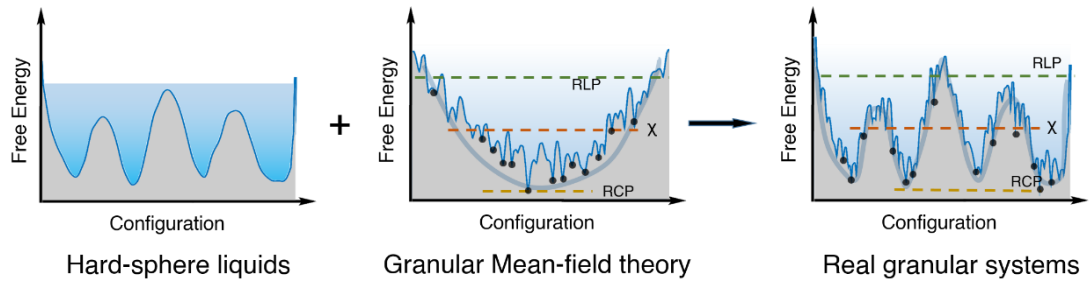

FIG. 4. Schematic illustration of free energy landscapes from both the mean-field perspective and real granular systems. Left: The energy landscape of a frictionless hard-sphere liquid is characterized by a complex multi-basin structure. Middle: The mean-field landscape confines all states to a single rugged basin. Each local minimum within this basin corresponds to a mechanically stable configuration, separated by energy barriers. Here, friction affects the density of states (DOS) and decides the threshold height of these minima (defining the RLP state), while compactivity governs their overall distribution. Right: In real granular systems, the energy landscape combines a disordered hard-sphere structure with intricate ruggedness and detailed features within individual basins.

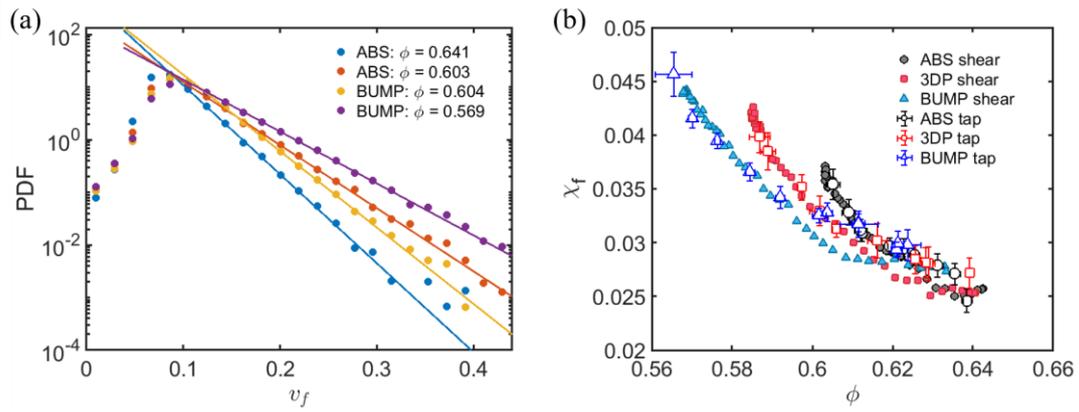

FIG. 5. (a) Probability distribution function (PDF) of the tetrahedron free volume $v_f$ for four representative packings by shearing. The solid lines represent fittings for exponential tails. (b) Compactivity $\chi_f$ as a function of $\phi$ for the three systems calculated via the slopes of PDFs, as shown in panel (a).

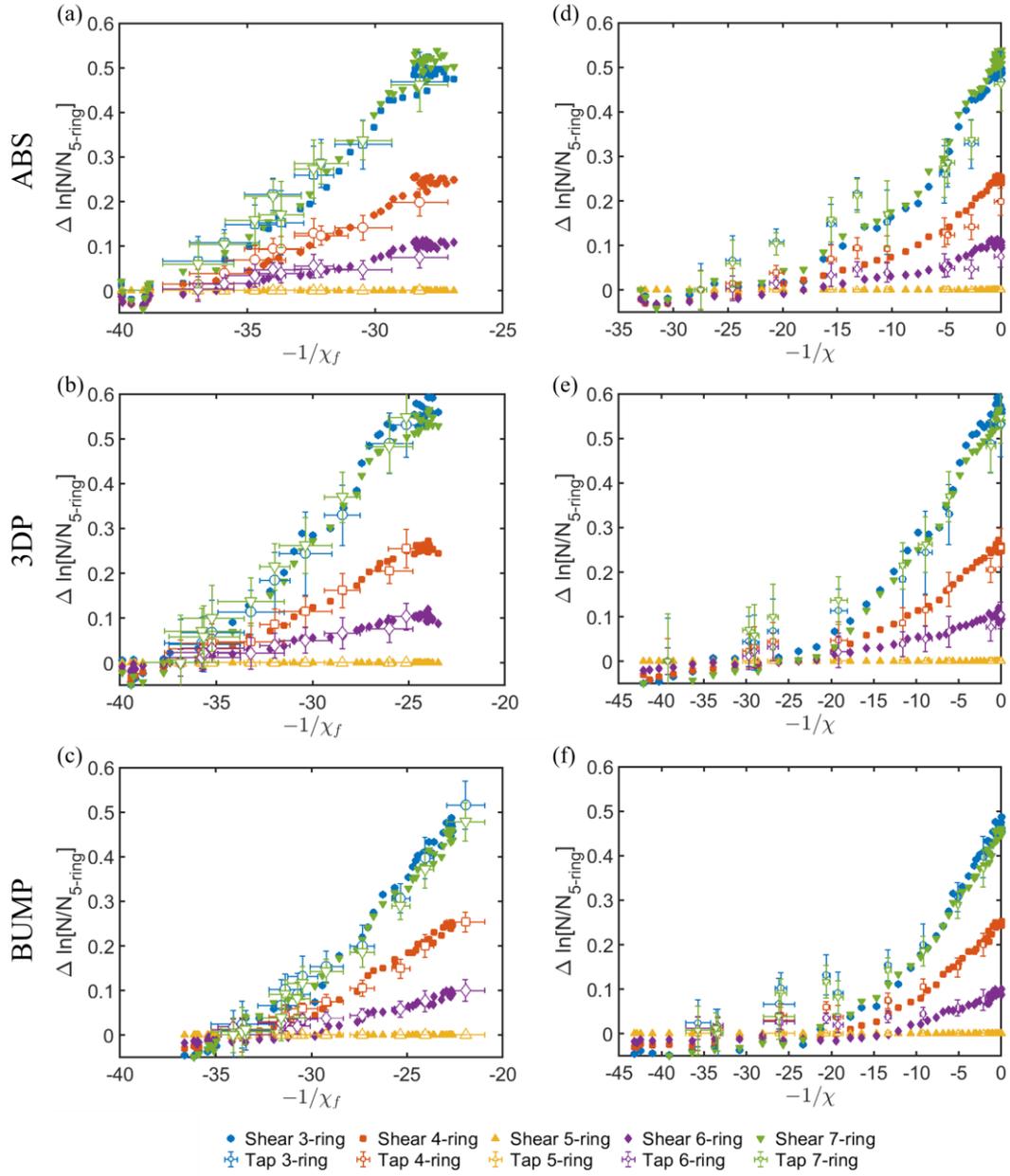

FIG. 6. $\Delta \ln(N_{N-ring}/N_{5-ring})$ as a function of (a)–(c) $-1/\chi_f$ and (d)–(f) $-1/\chi$. Panels (a) and (d), (b) and (e), and (c) and (f) correspond to ABS, 3DP, and BUMP particle systems, respectively. The legends for all panels are displayed at the bottom of the figure. All data were vertically shifted for clarity and comparison. The slopes in panels (a)–(c) correspond to the excitation volumes of the N-ring structures.

TABLE I. Excitation volume for N-ring structures(unit: $v_g$)

|        | ABS   |       | 3DP   |       | BUMP  |       |
|--------|-------|-------|-------|-------|-------|-------|
|        | tap   | shear | tap   | shear | tap   | shear |
| 3-ring | 0.043 | 0.042 | 0.042 | 0.040 | 0.040 | 0.039 |
| 4-ring | 0.021 | 0.021 | 0.020 | 0.020 | 0.020 | 0.021 |
| 6-ring | 0.007 | 0.009 | 0.007 | 0.008 | 0.007 | 0.008 |
| 7-ring | 0.044 | 0.042 | 0.044 | 0.040 | 0.039 | 0.039 |

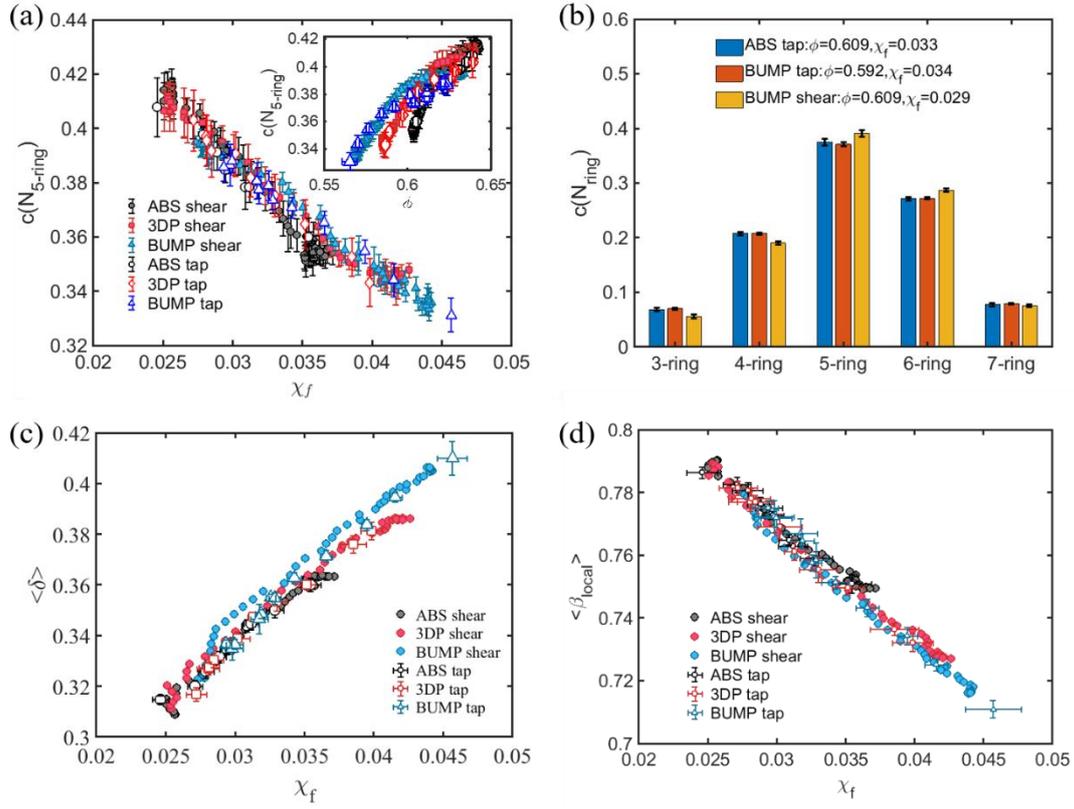

FIG. 7. (a) concentration of 5-ring structures $c(N_{5-ring})$ as a function of $\phi$ for different systems. (b) Concentration of 3 to 5-ring structures for three representative packings. (c) Average polytetrahedral order parameter $\langle \delta \rangle$ and (d) average anisotropy index $\langle \beta_{local} \rangle$ as a function of $\chi_f$.

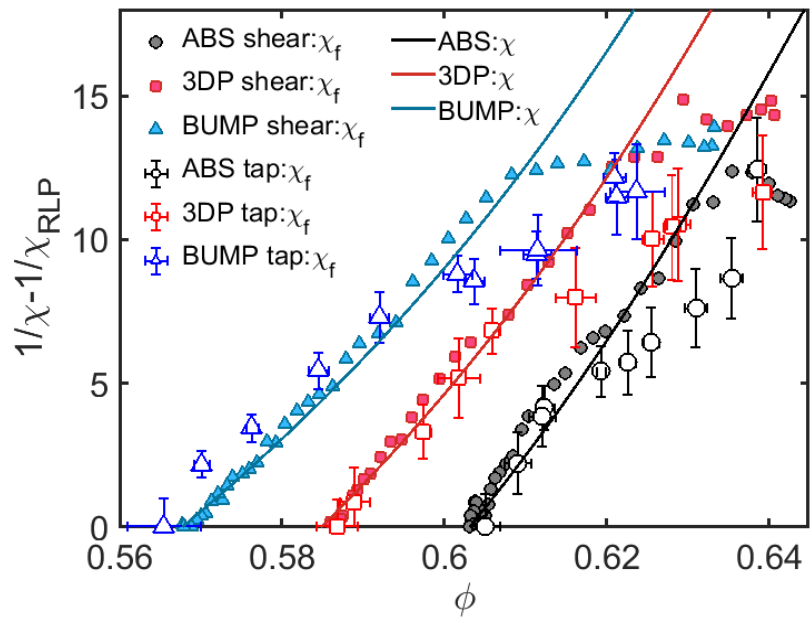

FIG. 8. Comparison of two effective temperatures, $\chi$ and $\chi_f$.